\documentclass[pra,notitlepage]{revtex4-1}

 \def\ep{{\epsilon}}

 \def\frac#1#2{{#1\over #2}}
\def\Tr{\text{Tr}}

 \def\L{{\Lambda}}

 \def\Re{{\rm Re}}
 \def\Im{{\rm Im}}
 
\def\be{\begin{equation}}
\def\ee{\end{equation}}
\def\bea{\begin{eqnarray}}
\def\eea{\end{eqnarray}}
\newcommand{\ket}[1]{\left| #1 \right>}

 \def\no{\nonumber \\}

 \def\la{\langle}
 \def\lb{\rangle}
 \def\ep{\epsilon}

\def\ba{\begin{eqnarray}}
\def\ea{\end{eqnarray}}

\usepackage{amssymb,amsmath,amstext}                
\usepackage{graphicx}                                               
\usepackage{color}                                                     
\usepackage{bm}                                                        
\usepackage{appendix}                                              
\usepackage[utf8]{inputenc}
\usepackage{bbold}
\usepackage{bbm}
\usepackage{ulem}
\usepackage{latexsym}
\usepackage[colorlinks=true,citecolor=blue,linkcolor=magenta]{hyperref}

\begin{document}
\begin{flushright}                                      
NORDITA-2016-56
\end{flushright} 
\title{Quantum dimensions from local operator excitations in the Ising model \vspace{0.6cm}}
\author{Pawe{\l}  Caputa$^{1}$ and Marek M. Rams$^2$ \vspace{0.6cm} }

\affiliation{$^1$Nordita, KTH Royal Institute of Technology and Stockholm University, Roslagstullsbacken 23, SE-106 91 Stockholm, Sweden\\
$^2$ Institute of Physics, Jagiellonian University,  \L{}ojasiewicza 11, 30-348 Krak\'ow, Poland%
\vspace{0.6cm}}%
\begin{abstract}
We compare the time evolution of entanglement measures after local operator excitation in the critical Ising model with predictions from conformal field theory. For the spin operator and its descendants we find that R\'enyi entropies of a block of spins increase by a constant that matches the logarithm of the quantum dimension of the conformal family. However, for the energy operator we find a small constant contribution that differs from the conformal field theory answer equal to zero. We argue that the mismatch is caused by the subtleties in the identification between the local operators in conformal field theory and their lattice counterpart. Our results indicate that evolution of entanglement measures in locally excited states not only constraints this identification, but also can be used to  
extract non-trivial data about the conformal field theory that governs the critical point. We generalize our analysis to the Ising model away from the critical point, states with multiple local excitations, as well as the evolution of the relative entropy after local operator excitation and discuss universal features that emerge from numerics.
\end{abstract}
\maketitle%

\section{Introduction}
The physics of many 1+1 dimensional lattice models at criticality is captured in the continuum limit by two-dimensional conformal field theory (2d CFT). As a result, one can numerically extract the conformal data, for instance scaling dimensions of the primary operators from the scaling of the two-point correlation functions in the critical lattice model, see \cite{DiF} for the standard reference. On the other hand, measures of entanglement proved to be useful quantities for exploring the critical points of many-body systems.  For example, in the ground state of a critical chain, the entanglement entropy of a block of spins has a universal logarithmic scaling with the size of the block that is proportional to the central charge \cite{Calabrese:2004eu}. This provides an efficient numerical way to obtain the central charge of the CFT that governs the critical point, see Refs.~\cite{Calabrese:2016xau} for review. It is then natural to ask if entanglement measures in excited states can be used to extract more CFT data, such as for instance the modular S or T matrices or quantum dimensions, numerically. 

In rational CFTs local elementary excitations are catalogued into finite number of conformal families containing primary operators and their descendants. The simplest excited states can then be obtained by inserting local CFT operators at some spatial points. In such states, one can study how the local operator changes the structure of entanglement in the ground state. More precisely, it is possible to compute the time evolution of the change in R\'enyi entropies for a reduced density matrix of a single interval due to the operator insertion \cite{REEQD} --  see Ref.~\cite{Nozaki:2014uaa,LEx} for some results in various CFT setups and Ref.~\cite{Global} for entanglement in a related class of globally excited states. In 2d CFT this analysis can be preformed analytically and R\'enyi entropies detect an increase in entanglement equal to the logarithm of the quantum dimension of the conformal family \cite{He:2014mwa,Caputa:2015tua,Chen:2015usa}. 

Having such a clear and elegant prediction from the CFT, it is then natural to wonder if and how the logarithms of quantum dimensions are reproduced on the lattice. In this article we initiate such program for the simplest case of the critical Ising chain. The advantage of the Ising model is that it is exactly solvable and the action of a family of local operators can be efficiently simulated for large system sizes. The main subtle point of this analysis is the identification between the CFT operators and their lattice counterpart. In fact, there are only few models where such map is well established -- see for instance the discussion in \cite{Mong:2014ova} -- and a general belief is that a given lattice operator corresponds in the continuum to a primary operator plus its descendants. In the case of the two-point functions these extra contributions from descendants lead to corrections that are suppressed as higher powers with the distance. In this work, given a well established identification for the Ising model operators, we will be able to check the contribution form this non-unique identification to the physics of entanglement propagation. We will see that for truly local operators on the lattice -- like the Ising spin -- we recover the CFT answer, but for operators with non-local support -- like Ising energy -- the subleading contributions modify the leading answer and lead to a mismatch.

The computations of R\'enyi entropies in CFT are done using the replica method that, for excited states, boils down to calculation of correlation functions on complicated Riemann surfaces. Even for states locally excited by more than a single operator such objects are notoriously difficult to compute analytically and features of entanglement measures in this class of states remain unexplored. Similarly, measures of distance between quantum states like, e.g., relative entropy for locally excited states require the access to higher-point correlators \cite{Lashkari:2015dia}. In this work, we will further explore the Ising model to shed a new light in these directions by numerically performing the time evolution of the relative entropy, as well as R\'enyi entropies in more general states excited by multiple local operators.

This paper is organized as follows. In section \ref{sec:CFT} we summarize the relevant results from the two dimensional CFT. In section \ref{sec:num} we present our main numerical results for the evolution of entanglement in the critical Ising model in states excited by single local operators. In section \ref{sec:gen} we consider the evolution away from the critical point, as well as more general operator excitations. In section \ref{sec:rel} we present the evolution of the relative entropy after local operator excitation in this model. Finally, we conclude and present the details of our numerical approach in the Appendix \ref{Numer}.

\begin{figure*}[t]
\begin{center}
\includegraphics[width=90mm]{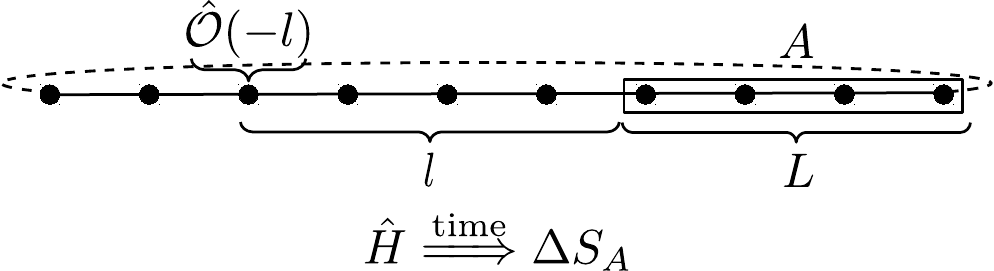}
\end{center}
\caption{Schematic illustration of the protocol. Local operator $\hat{\mathcal{O}}$ is inserted at a distance $l$ from the block $A$ of $L$ spins. We then calculate the change of entropy of the block  $\Delta S_A$  resulting from the insertion as a function of time.}
\label{fig:schematic}
\end{figure*}

\section{CFT Results} \label{sec:CFT}
 In this section we briefly review the existing results for the evolution of R\'enyi entropies in locally excited states in 2d CFTs and then show some details of the computation for the Ising CFT. Finally, we discuss the minor modifications that appear for the CFTs on the cylinder that, in the following sections, we will be comparing to numerics from the periodic chain.
 
Let us start with a 2d CFT on the real line and a state excited by a local operator $\hat{\mathcal{O}}(-l)$ at distance $l$ from the entangling interval $A\in [0,L]$, as presented pictorially on Fig.~\ref{fig:schematic}. The density matrix is given by 
\bea
\hat{\rho}(t)&=&{\mathcal N}\cdot e^{-i\hat{H}t}e^{-\ep \hat{H}}\hat{\mathcal{O}}(0,-l)|0\lb\la 0|\hat{\mathcal{O}}^{\dagger}(0,-l)e^{-\ep \hat{H}}e^{i\hat{H}t} \no
&\equiv& {\mathcal N}\cdot \hat{\mathcal{O}}(w_2,\bar{w}_2)|0\lb\la 0|
\hat{\mathcal{O}}^{\dagger}(w_1,\bar{w}_1),  \label{denmat}
\eea
where the insertion points of the local operators are defined as
\bea
&& w_1=i(\epsilon -it)-l, \ \ w_2 = -i(\epsilon+it)-l, \nonumber\\
&& \bar{w}_1=-i(\ep-it)-l,\ \ \bar{w}_2=i(\epsilon+it)-l. \label{points}
\eea
The factor of $\epsilon$ is the UV regulator for the local operators and we take $\epsilon\to 0$ at the end of the computation. The normalization $\mathcal{N}$ ensures that the trace of the density matrix is equal to 1.\\
Using the replica trick, we can compute how a family of R\'enyi entropies $S_A^{(n)}$, indexed by integer $n$, changes due to the local operator insertion. The answer to this question is expressed in terms of the logarithm of the ratio \cite{REEQD}
\bea
\Delta S_A^{(n)}\equiv\frac{1}{1-n}\log\left[\frac{\langle {\hat{\cal O}}(w_1, \bar{w}_1){\hat{\cal O}}^{\dagger}(w_2, \bar{w}_2)\cdots {\hat{\cal O}}^{\dagger}(w_{2n}, \bar{w}_{2n})\rangle_{\Sigma_n}}{\left(\langle {\hat{\cal O}}^{\dagger}(w_1, \bar{w}_1){\hat{\cal O}}(w_2, \bar{w}_2)\rangle_{\Sigma_1}\right)^n}\right],
\label{eq:Delta}
\eea
where $\Delta S^{(n)}_A$ is the difference between the $n$-th R\'enyi entropy of  a subregion $A$ computed in excited state by the local operator and the vacuum. The correlator in the numerator is computed on the $n$-sheeted surface $\Sigma_n$ with cuts on each copy corresponding to interval $A$, and the two-point function in the denominator is on a single sheet $\Sigma_1$ with an interval cut $A$. For a detailed derivation and further illustrative explanations see \cite{REEQD}.

In 2d CFT, one can apply a conformal map from $\Sigma_n$ to a complex plane and evaluate the correlators explicitly. It turns out that the answer is universal and the increase in the R\'enyi entanglement entropies is equal to a constant that is the same for all the members of a conformal family, i.e. primary operators and their descendants \cite{Caputa:2015tua,Chen:2015usa}. In rational CFTs this constant is equal to the logarithm of the quantum dimension of the local operator \cite{He:2014mwa} which is defined as 
\be
d_a=\frac{S_{0a}}{S_{00}},\label{qdim}
\ee  
where $S_{ij}$ denotes the elements of the modular S-matrix of the CFT, see e.g. Ref.~\cite{DiF}.

In this work we focus on the 2d Ising model which is the $(4,3)$ minimal model with three primary operators: the identity $\mathbf{1}$, the energy $\varepsilon$ with conformal dimensions $(h_\varepsilon,\bar{h}_\varepsilon)=\left(\frac{1}{2},\frac{1}{2}\right)$  $(\Delta_\varepsilon=h_\varepsilon+\bar{h}_\varepsilon=1)$ and the spin $\sigma$ with $(h_\sigma,\bar{h}_\sigma)=\left(\frac{1}{16},\frac{1}{16}\right)$ $(\Delta_\sigma=h_\sigma+\bar{h}_\sigma=\frac{1}{8})$, where $\Delta_{\sigma(\varepsilon)}$ marks the total scaling dimensions.\\
The modular S-matrix of the Ising model is given by
\bea
S=\frac{1}{2}\left(
\begin{array}{ccc}
 1 & 1 & \sqrt{2} \\
 1 & 1 & -\sqrt{2} \\
 \sqrt{2} & -\sqrt{2} & 0 \\
\end{array}
\right),\eea
so the three quantum dimensions \eqref{qdim} are
\be
d_{\mathbf{1}}=d_\varepsilon=1,\qquad d_\sigma=\sqrt{2}.
\ee
This way, at criticality, only excitations with primary $\sigma$ can non-trivially change the entanglement in the vacuum state, and for all the R\'enyi entropies we have \cite{He:2014mwa}
\be
\Delta S^{(n)}_A=\log \sqrt{2}.
\ee
The standard, chiral (anti-chiral) descendants are obtained by either acting on the primary operators with chiral (anti-chiral) derivatives $\partial_z$ ($\bar{\partial}_{\bar{z}}$) or taking the operator product expansion (OPE) of the primary operators with the energy momentum tensor. Such descendants increase the entropies by the same amount as the primaries. If we however act with the linear combination of the two derivatives, there is an additional contribution to the entropy equal to $\log2$ \cite{Chen:2015usa}. We will see this in case of the spatial derivative $\partial_x=\partial_z+\bar{\partial}_{\bar{z}}$ acting on $\sigma(z,\bar{z})$, which we consider in Sec.~\ref{sec:num} C.
\begin{figure}[t]
\begin{center}
\includegraphics[width=70mm]{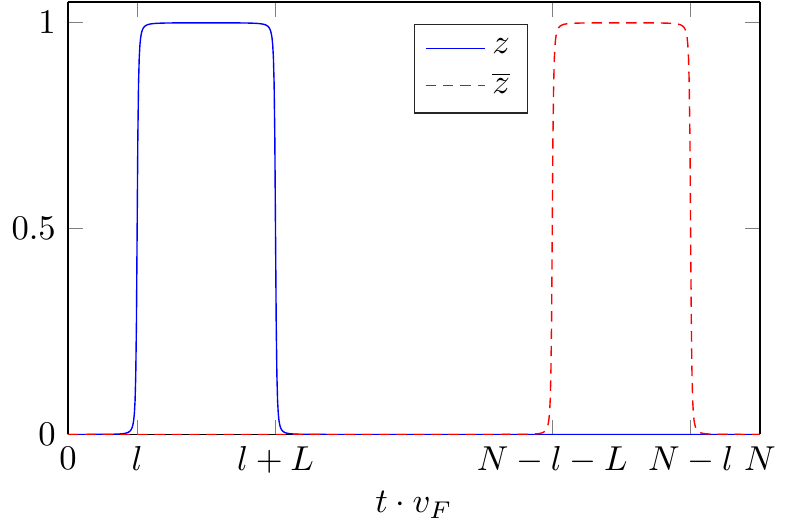}
\end{center}
\caption{Evolution of the cross-ratios on the cylinder for a single period of time for small but non-zero $\epsilon$.}
\label{fig:CR1}
\end{figure}

In order to compare the CFT results with numerics we have to take into account finite size of the system, $N$. In the CFT computation it enters through the invariant cross-ratios. Let us, for simplicity, consider the change in the second Renyi entropy $\Delta S^{(2)}_A$ that requires the correlator on two cylinders. The correlators entering \eqref{eq:Delta} are computed using a composition of the conformal map $w(x)=\exp\left(-\frac{2\pi i}{N}x\right)$ from each cylinder to the plane with a cut and the uniformization map $z^2(w)=(w-1)/(w-w(L))$. After some standard CFT manipulations, the change in the entropy can be written as 
\be
\Delta S^{(2)}_A=-\log\left((z(1-z))^{2 h_\mathcal{O}} (\overline z(1-\overline z))^{2 h_\mathcal{O}}\mathcal{G}^{\mathcal{O}\mathcal{O}}_{\mathcal{O}\mathcal{O}}(z,\bar{z})\right),\label{SecRen}
\ee
where $\mathcal{G}^{\mathcal{O}\mathcal{O}}_{\mathcal{O}\mathcal{O}}$ is the canonical 4-point function on the complex plane with operators $\mathcal{O}$ inserted at $(0,z,1,\infty)$ and the cross-ratios are defined as
\be
z=\frac{z_{12}z_{34}}{z_{13}z_{24}},\qquad \bar{z}=\frac{\bar{z}_{12}\bar{z}_{34}}{\bar{z}_{13}\bar{z}_{24}},
\ee
with $z_i\equiv z(w_i)$, $z_{ij}=z_i-z_j$ and similarly for $\bar{z}$. From the conformal map, we can also show that $z_3=-z_1$ and $z_4=-z_2$ (similarly for $\bar{z}$). 

Once we plug the insertion points of the operators \eqref{points}, in the $\epsilon\to 0$ limit, the cross-ratios become periodic functions of time, as shown on Fig.~\ref{fig:CR1}.  The difference with the CFT on the infinite line is that in one cycle of time equal to $N$, both $z$ and $\bar{z}$ reach their maximal value of $1$. More precisely, $z\sim1$ in the time $[t_{i},t_o]=[l+\alpha N,l+L+\alpha N]$ and zero outside, whereas  $\bar{z}\sim1$ inside $[\bar{t}_{i},\bar{t}_o]=[N-(l+L)+\alpha N,N-l+\alpha N]$ and zero outside, for integer period $\alpha$.
In order to extract the increase in the second R\'enyi entropy analytically in the $\epsilon\to 0$ limit, we simply take the limit of $(z,\bar{z})\to(1,0)$ or  $(z,\bar{z})\to(0,1)$ in \eqref{SecRen} in the appropriate time intervals.

Let us focus on the explicit example of the Ising model. For the $\sigma$ operator, from the fusion rule $\sigma\times \sigma=1+\varepsilon$, the correlator can be decomposed as \cite{DiF}
\be
\mathcal{G}^{\sigma\sigma}_{\sigma\sigma}(z,\bar{z})=\left(C^1_{\sigma\sigma}\right)^2\mathcal{F}^{\sigma\sigma}_{\sigma\sigma}(1|z)\mathcal{F}^{\sigma\sigma}_{\sigma\sigma}(1|\bar{z})+\left(C^\varepsilon_{\sigma\sigma}\right)^2\mathcal{F}^{\sigma\sigma}_{\sigma\sigma}(\varepsilon|z)\mathcal{F}^{\sigma\sigma}_{\sigma\sigma}(\varepsilon|\bar{z}),
\ee
where the three-point coefficients $C^1_{\sigma\sigma}=1$ and $C^\varepsilon_{\sigma\sigma}=\frac{1}{2}$, as well as the conformal blocks
\bea
\mathcal{F}^{\sigma\sigma}_{\sigma\sigma}(1|z)=\frac{1}{\sqrt{2}}\frac{\sqrt{1+\sqrt{1-z}}}{(z(1-z))^{\frac{1}{8}}},\qquad\mathcal{F}^{\sigma\sigma}_{\sigma\sigma}(\varepsilon|z)=\sqrt{2}\frac{\sqrt{1-\sqrt{1-z}}}{(z(1-z))^{\frac{1}{8}}}.
\eea
We can then check that that the non-zero contribution in the two time intervals comes from the identity block and is equal to $\Delta S^{(n)}_A=\log \sqrt{2}$, in accordance with the modular S-matrix elements. This behavior is naturally explained from the quasi-particle picture where left and right moving sets of entangled quasi-particles propagate from the insertion point of the operator and, on the circle, there are two time intervals where either left or right particles are inside the entangling interval $A$.

On the other hand, for the $\varepsilon$ excitation, using the fusion $\varepsilon\times\varepsilon=1$, we can write the correlator as
\be
\mathcal{G}^{\varepsilon\varepsilon}_{\varepsilon\varepsilon}(z,\bar{z})=\left(C^1_{\varepsilon\varepsilon}\right)^2\mathcal{F}^{\varepsilon\varepsilon}_{\varepsilon\varepsilon}(1|z)\mathcal{F}^{\varepsilon\varepsilon}_{\varepsilon\varepsilon}(1|\bar{z}),
\ee
with $C^1_{\varepsilon\varepsilon}=1$ and conformal block
\be
\mathcal{F}^{\varepsilon\varepsilon}_{\varepsilon\varepsilon}(1|z)=\frac{1-z+z^2}{z(1-z)}. 
\ee
Clearly, inserting the cross-ratios and taking $\epsilon\to 0$ yields $\Delta S^{(n)}_A=0$ for all times. This suggests that the quasiparticles produced by $\varepsilon$ are in a product state. 

After this short review, below we perform the numerical analysis and compare how the CFT predictions are reproduced on the discrete chain.

\section{Single excitations in Ising spin chain} \label{sec:num}
We consider quantum Ising model in a transverse magnetic field on a 1d chain of $N$ spins-$1/2$ described by the Hamiltonian
\be
\hat H=-\sum_{n=1}^N\left[\hat \sigma^x_n \hat \sigma^x_{n+1}+g \hat \sigma^z_n\right],
\label{eq:Spin_Ising}
\ee
where $\hat \sigma^{x,z}_n$ are the standard Pauli matrices acting on $n$-th spin and we assume periodic boundary conditions $\vec{\hat \sigma}_1 = \vec{\hat \sigma}_{N+1}$.
The model is critical for $g=\pm1$ and unless stated otherwise, in this article we  set $g=1$. 

The model can be mapped onto the free fermion system using the Jordan-Wigner transformation and our numerical results are performed in such a setup. We refer to the Appendix \ref{Numer} for details. After the mapping, the Hamiltonian can be diagonalized as

\be
\hat H=\sum_{k} \epsilon_k \left(\hat \gamma_k^\dagger \hat \gamma_k - \frac12\right),
\label{eq:H_Ising}
\ee
where $\hat \gamma_k$ are the fermionic annihilation operators, and we refer to  Appendix \ref{Numer} for some technical subtleties related with the boundary conditions.

The ground state of $\hat H$ is the vacuum state annihilated by all $\hat \gamma_k$ and the dispersion relation reads
\be
\epsilon_k = 2 \sqrt{(g-\cos k)^2 + \sin^2 k}\, ,
\label{eq:epsilon_k}
\ee
where the quasi-momenta $k$ take discrete values $\in [-\pi , \pi]$.
Even at the critical point, for $g=1$, the dispersion relation is not strictly linear for all values of $k\in [-\pi , \pi]$ due to the discrete nature of the system, approaching linear behaviour only in the limit of long-wavelength $|k| \ll 1$. As a result, the velocity of the quasiparticles depends on the momentum $k$, especially for larger values of $|k|$ -- in contrast with what is the case for CFT. In order to account for that, as well as for comparisons with CFT, we simultaneously consider the model with linearized dispersion relation fixing the velocity of quasiparticles, 
\be
\hat H_{lin}=\sum_{k} \epsilon^{lin}_k \left(\hat \gamma_k^\dagger \hat  \gamma_k - \frac12\right),
\ee
where $\epsilon^{lin}_k = v_F |k|$ and  $\hat \gamma_k$ are the annihilation operators diagonalizing Ising Hamiltonian in Eq.~\eqref{eq:H_Ising}. In our conventions, $v_F = 2$ is the velocity of quasiparticles at the critical point for that Hamiltonian in the long-wavelength limit of $k\to 0$, but we will present our plots appropriately rescaled for comparisons with CFT.

Next, we study the evolution of R\'enyi entanglement entropies in states locally excited by operators on the lattice. More precisely, we excite the ground state of the critical Ising model with local operators $\mathcal{O}(n)$, which can have support on more then one lattice site, and perform a unitary time evolution. We then numerically calculate the entropy of a block $A$ of $L$ consecutive spins at a distance $l$ from the excitation for different times and subtract from it the entropy of the block with no excitation. This setup is schematically illustrated in Fig.~\ref{fig:schematic}. 

The lattice operators that have $\sigma$ and $\epsilon$ fields as their leading contributions in the continuum are \cite{Stojevic:2014zta}
\bea
\sigma(n)&=&\hat \sigma^x_n,\\
\varepsilon(n)&=&\hat \sigma^x_n \hat \sigma^x_{n+1}- \hat \sigma^z_n.
\eea
Note the opposite sign of the $\sigma^z_n$ term to \eqref{eq:Spin_Ising}.
To fix the $\varepsilon$ operator on the lattice we can use the fact that it should take the Hamiltonian away from criticality (mass term in the free fermions language) and, as in the continuum CFT, should be odd under the Kramers-Wannier duality that on the lattice interchanges both terms.\\

Moreover, to confront the CFT predictions for other members of a given conformal family, we also consider the simplest descendant, namely the spatial discrete derivative of the $\sigma$ field
\be
d\sigma(n)=\hat \sigma^x_{n+1} - \hat \sigma^x_n.
\ee 
that corresponds to descendant $\partial_{x}\sigma(z,\bar{z})$.\\
In the following sections, we discuss the results for classes of states locally excited by these three different operators.

\subsection{$\sigma(n)$ excitation}
\begin{figure*}[t]
\begin{center}
\includegraphics[width=\linewidth]{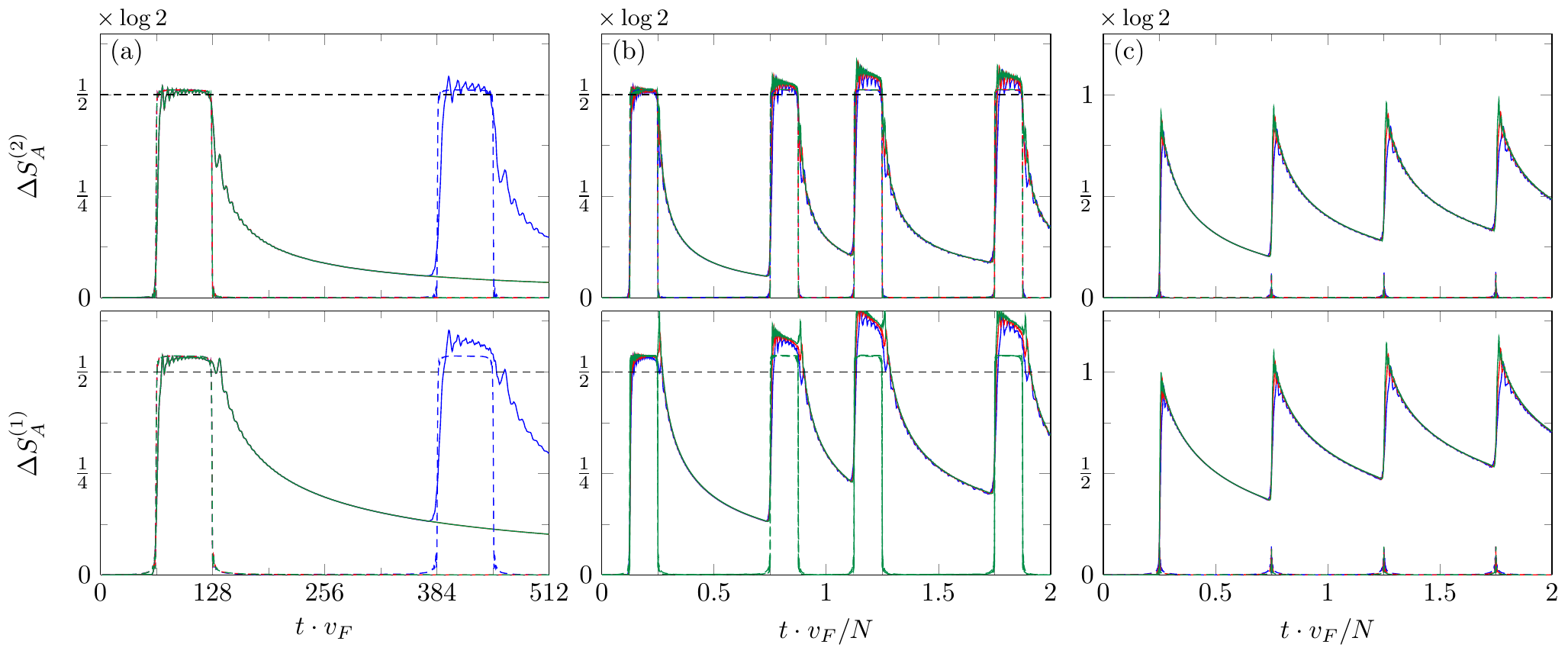}
\end{center}
\caption{Evolution after excitation by $\sigma(n)$. Results for Ising Hamiltonian (solid lines) and linearized Ising Hamiltonian (dashed lines). Different system sizes $N=512$ (blue), $N=1024$ (red) and $N=2048$ (green). (a) Fixed block size $L = 64$ and distance $l = 64$.  (b) Distance and block size as a fraction of the system size, $L=l=N/8$. (c) Excitation in the same distance from both ends of the block; Block size as a fraction of the system size $L = N/2+1$, $l=N/4$. See text for discussion.}
\label{fig:x}
\end{figure*}

The results for $\sigma(n)$ excitation are collected in Fig.~\ref{fig:x}. Column (a) shows the change in entropy for fixed $L$ and $l$. We observe the expected plateau appearing at a time when $t \cdot v_F \simeq l$ -- and subsequent plateaus when the signal enters the block $A$ from the opposite side or after making some number of circle around the chain. The plateau is however oscillating slightly and then vanishing in a long tail for $t \cdot v_F > l+L$. The mean values in the first  plateau are $\Delta S_A^{(2)} \approx 0.51 \log 2$ and $\Delta S_A^{(1)} \approx 0.54 \log 2$, close to the expected value of $\log \sqrt{2}$. Both the oscillations and the tail visible for the first plateau are mostly independent of the system size.
When we use linearized Ising Hamiltonian $\hat H_{lin}$ (dashed lines) in place of Ising Hamiltonian $\hat H$ (solid lines) both the oscillations and the tails disappear and fully periodic structure consistent with the CFT prediction is recovered. This supports the natural interpretation that the tail is related to different velocities of excited quasiparticles which naturally appear for local Hamiltonian on a chain and in this case are smaller then $v_F$. Notice also that the dashed line has smooth edges, what can be interpreted as manifestation of the nonzero $\epsilon$ on the lattice, compare with Fig.~\ref{fig:CR1}. Finally, for $\hat H$, the subsequent plateaus  appearing for longer times are visibly shifted up, as the signal is slowly dissipating.
 
Column (b) shows the change of entropy when both $l$ and $L$ are proportional to the system size $N$ and the time is properly rescaled.  This validates that the obtained value of $\Delta S_A$ is independent of the block size -- in contrast to the subtracted background entropy of the block without excitation which in this case grows logarithmically with $L\sim N$. Additionally, we observe that after the rescaling the tails for different $L\sim N$ are collapsing  -- at least up to corrections which are not visible in this scale and for short enough times -- which suggests that the characteristic time scale at which the tails are disappearing is proportional to the block size.

Finally, in column (c) we place the excitation symmetrically with respect to the block, so that the quasiparticle with the same absolute momentum traveling left and right should be entering the block at the same time. Even in this setup $\Delta S_A$ acquires large non-zero value when the fastest quasiparticles traveling with $v_F$ reach block $A$ from both sides.  This shows that  the simplest interpretation valid for global (translationally invariant) quench that only the pairs of quasiparticles with opposite momenta $\pm k$ contribute to the entanglement is no longer valid for local excitation which breaks translational invariance. This signal almost disappears when $\hat H_{lin}$ is used and all the quasiparticles enter or leave the  block in the same instance of time. We refer for a consistent discussion in the context of local quenches to \cite{PeschelEislerLoc} and evolution of the negativity to \cite{Wen:2015qwa}.

\subsection{$\varepsilon(n)$ excitation}

\begin{figure*}[h]
\begin{center}
\includegraphics[width=\linewidth]{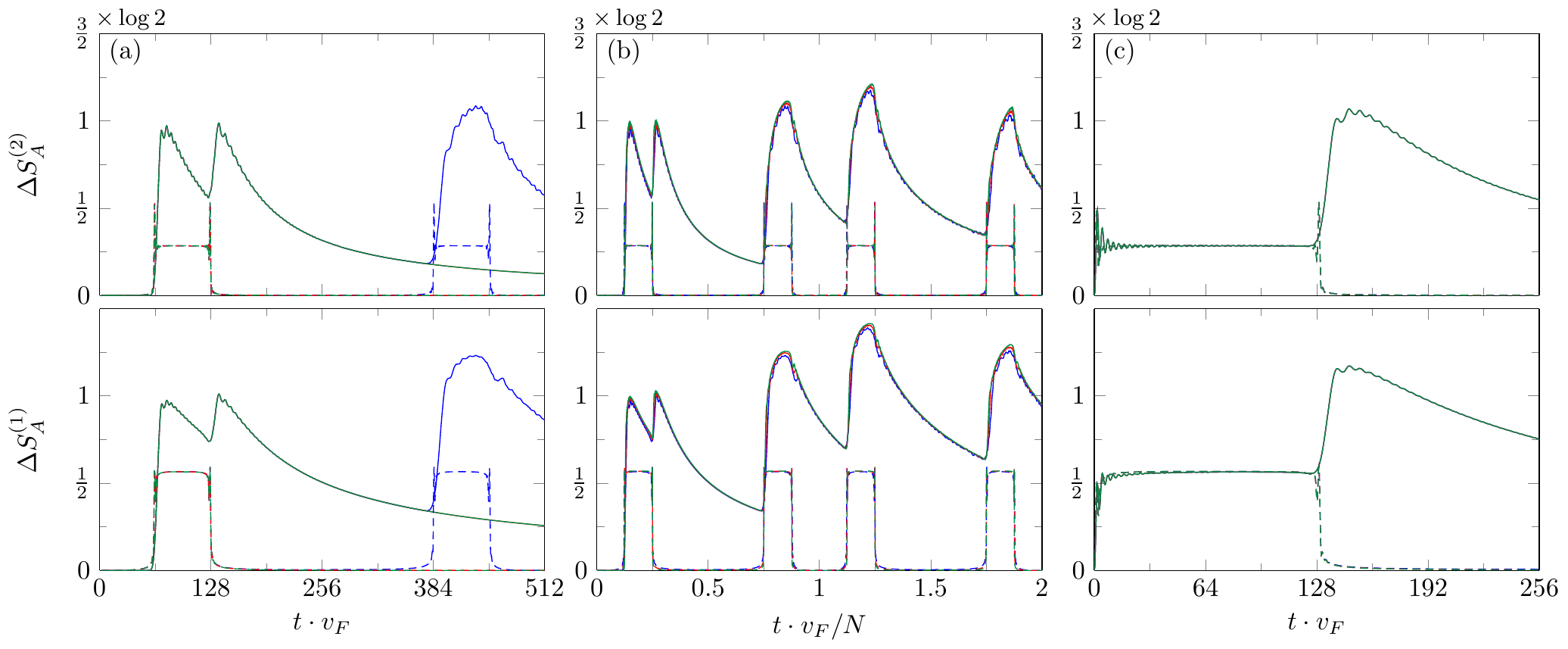}
\end{center}
\caption{Evolution after excitation by $\varepsilon(n)$. Results for Ising Hamiltonian (solid lines) and linearized Ising Hamiltonian  (dashed lines). System sizes $N=512$ (blue), $N=1024$ (red) and $N=2048$ (green).  (a) Fixed block size $L = 64$ and distance from the block $l = 64$;  (b) Block size and distance as a fraction of the system size, $L=l=N/8$. (c) Block next to the excitation with $l=1$ and the block size $L=128$. See text for discussion.}
\label{fig:e}
\end{figure*}

The results for $\varepsilon(n)$ excitation are presented in Fig.~\ref{fig:e}. For fixed $l$ and $L$ in column (a) we obtain non-zero signal with sharp peaks when the signal is first entering and leaving the block, which then disappears in a long tail.  When the evolution is governed by $\hat H_{lin}$ the tails are not present and we recover the periodic structure of plateaus, which however  acquire non-zero values of $\Delta S_A^{(2)} \approx 0.28 \log 2$, and $\Delta S^{(1)} \approx 0.56 \log 2$. Interestingly, in this case, there are still additional sharp peaks when the signal is entering or leaving the block, which are similar to the structure obtained from CFT with small but finite $\epsilon$. We are however not able to explain the value of $\Delta S_A$ in the plateaus as the logarithm of the quantum dimension of $\varepsilon$ that should be zero. We conclude that the sub-leading contributions to $\varepsilon(n)$  in the continuum turn out to be more relevant than for $\sigma(n)$. At the same time the observed results suggest that the quasiparticles with large $k$ significantly contribute to the observed value of entanglement, which might be related here with the fact that $\varepsilon(n)$ is supported on two lattice sites.

The stability of the obtained signal is validated in column (b) where we set $L$ and $l$ proportional to system size $N$, and after properly rescaling of the time we observe the collapse of $\Delta S_A$ for different values of $N$.

Finally, in (c) we show that the non-zero value of $\Delta S_A$ obtained in the plateaus for $\hat H_{lin}$ can be also recovered from the evolution governed by the original Ising Hamiltonian $\hat H$ when the block $A$ is placed just next to excitation.
This way difference in velocity of excited quasiparticles turn out to be unimportant for the first plateau as (almost) all right-moving quasiparticles are able to enter the block and the situation resembles that for $\hat H_{lin}$.  
This suggests that similar strategy can be used in general spin chains, which cannot be mapped onto  system of free fermions which we use to construct $\hat H_{lin}$. Such systems can be conveniently simulated using the toolbox of matrix product states (MPS) \cite{MPS_reviews}, where, for instance, specific algorithms to study the dynamics of localized excitations in infinite systems have been put forward \cite{MPS_comoving}.

\subsection{$\sigma(n+1)-\sigma(n)$ excitation}

\begin{figure*}[h]
\begin{center}
\includegraphics[width=\linewidth]{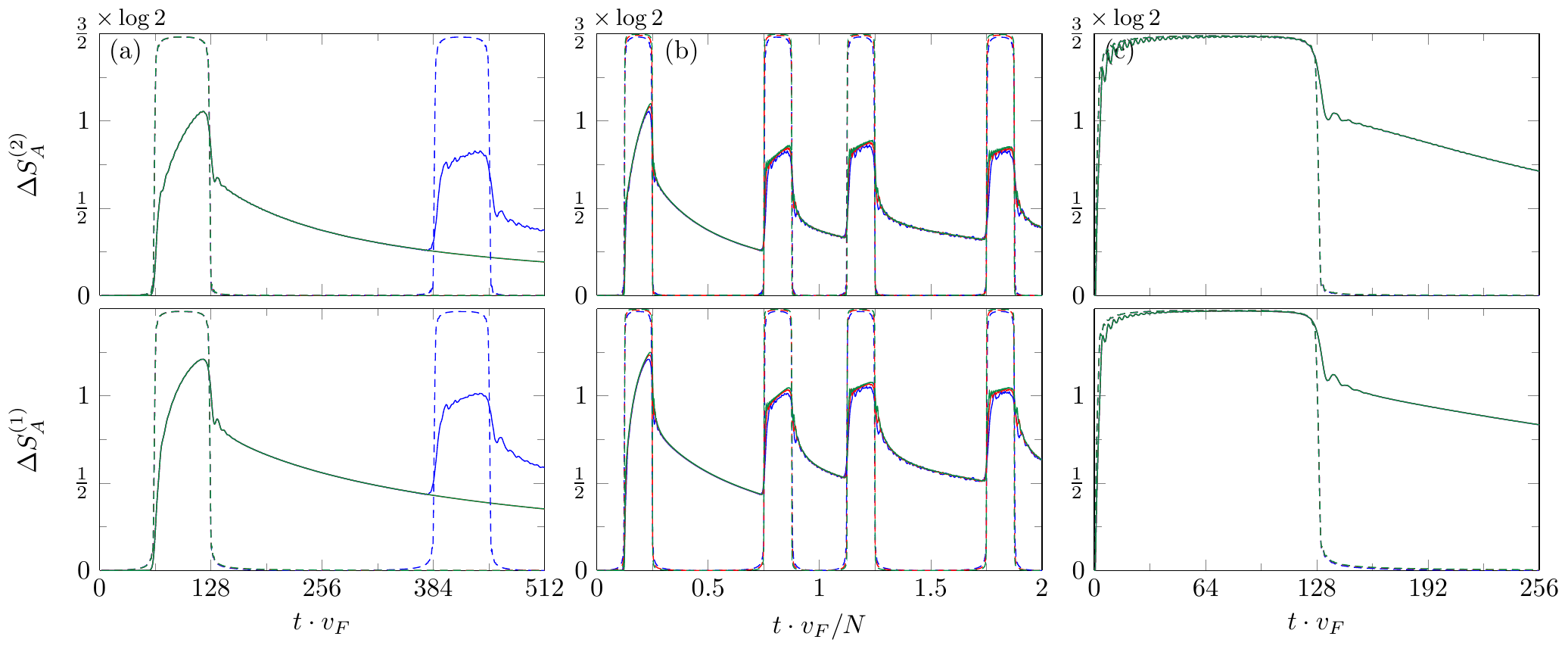}
\end{center}
\caption{Evolution after excitation by $\sigma(n+1)-\sigma(n)$. Results for Ising Hamiltonian (solid lines) and linearized Ising Hamiltonian  (dashed lines). System sizes $N=512$ (blue), $N=1024$ (red) and $N=2048$ (green).  (a) Fixed block size $L = 64$ and distance from the block $l = 64$;  (b) Block size and distance as a fraction of the system size, $L=l=N/8$. (c) Block next to the excitation with $l=1$ and the block size $L=128$. See text for discussion.}
\label{fig:dx}
\end{figure*}

The results for the discrete spatial derivative of $\sigma(n)$ are collected in Fig.~\ref{fig:dx}. We note that, on the contrary to the two previous cases, the operator $d\sigma(n) = \sigma(n+1)-\sigma(n)$ is not unitary, so even at $t=0$ it changes the entropy of a block supported on sites other then $n$ and $n+1$. We however see that this effect is relatively small and vanishing with increasing $l$.

In column (a), for fixed $l $ and $L$, we observe that $\Delta S_A$ does not form plateaus and then disappears in a long tails, which suggests that the slower quasiparticles with large $k$ significantly contribute to the observed signal.  This is further corroborated by using  $\hat H_{lin}$ which allows to recover the periodic structure of plateaus with the values at the peaks close to $\frac{3}{2}  \log 2$ predicted by CFT. Similar observation also hold in column (b) for $l$ and $L$ proportional to the system size $N$.
Finally, in column (c), similarly to the situation in the previous section, we observe that for local Ising Hamiltonian $\hat H$ we are able to recover the structure of the first plateau with the peak value  $\simeq 1.5  \log 2$  if the block $A$ is placed just next to the excitation and $l=1$.

\section{Non-critical evolution and general excitations} \label{sec:gen}
In this section we present numerical results that, in principle, could be reproduced from CFT, but in practice the analytical computations become very difficult. In such cases numerics is a great tool for understanding the phenomenology of entanglement evolution and we explore it below. Since we successfully recovered the CFT predictions for the spin $\sigma$ operator, we will mostly consider states excited by applying $\sigma^x$ to the lattice sites, but we believe that the universal features of our analysis remain valid for other conformal families.\\

\begin{figure}[h!]
\begin{center}
\includegraphics[width=14cm]{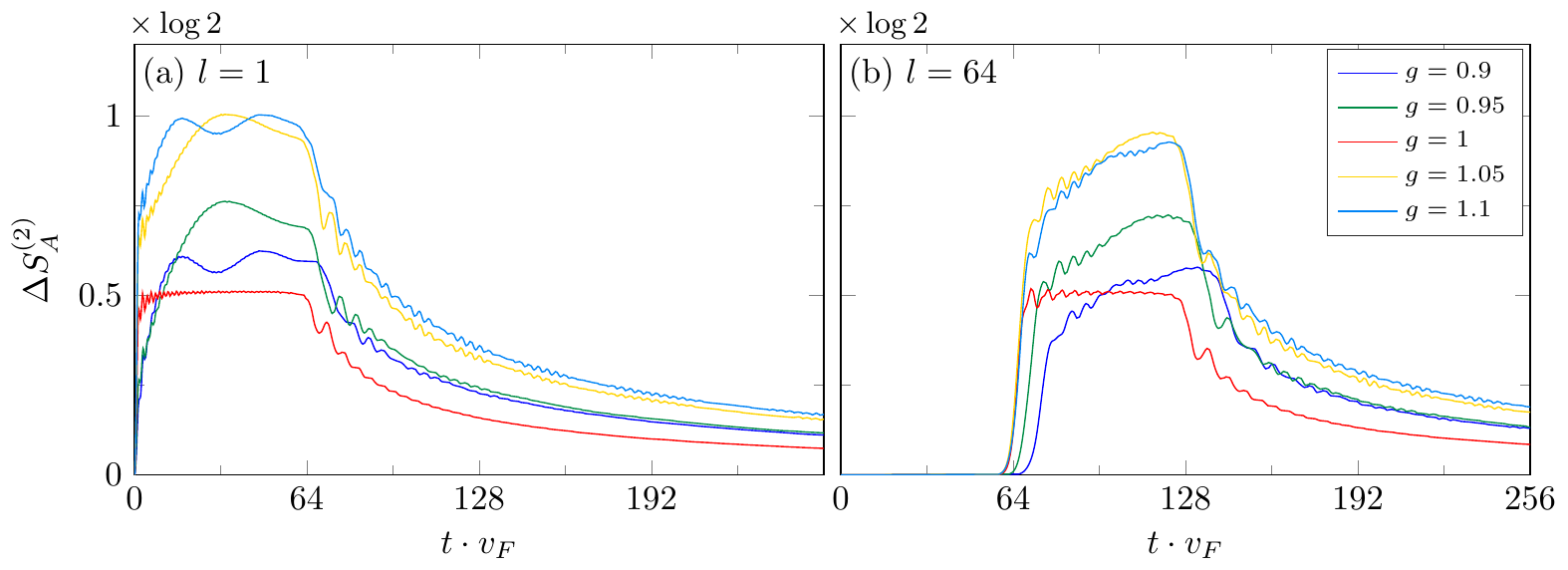}
\end{center}
\caption{Evolution of the change in the second R\'enyi entropy after $\sigma(n)$ excitation for different values of the magnetic field $g$. The distance of excitation from the block is (a) $l=1$ and (b)  $l=64$.  $L=64$, $N=512$.}
\label{Non-crit}
\end{figure}

We begin with evolution of entanglement of a block of spins after exciting the ground state by local operator but in the non-critical model. Having the diagonalized Ising Hamiltonian for any value of the parameters, we can study the evolution of the R\'enyi entropies away from the critical point. In Fig.~\ref{Non-crit} we plot the evolution of the second R\'enyi entropy after acting with $\sigma(n)$ for a few values of $g$ around the critical one. 

Clearly, in each case the entropy only changes after time of order $l$, however the clear plateau with the logarithm of the quantum dimension only appears at criticality. This, in principle, might serve as a proxy for seeing the critical point with local excitations but in general the value of the plateau might be a more complicated expression in terms of the quantum dimensions of the model. 

Figure~\ref{Non-crit} also indicates that the critical Hamiltonian leads to the smallest increase of entanglement of the block after we excite a vacuum by a local operator. Based on our limited analysis, it is not obvious if this lower bound is universal and satisfied by general families of Hamiltonians that have a critical point in their parameter space. It would be very interesting to provide further checks using other available methods -- like e.g. MPS -- in order to support this observation or possibly find counter-examples. \\

Next, by using the map to free fermions we can also study more general excited states by acting with multiple local operators on the critical chain. On the other hand, the computations in CFT using the replica trick become very cumbersome and were only done for two excitations in \cite{Tokiro}. This part will then serve as a collection of new predictions for the evolution of the R\'enyi entanglement entropies in (rational) CFTs.

\begin{figure}[h!]
\begin{center}
\includegraphics[width=14cm]{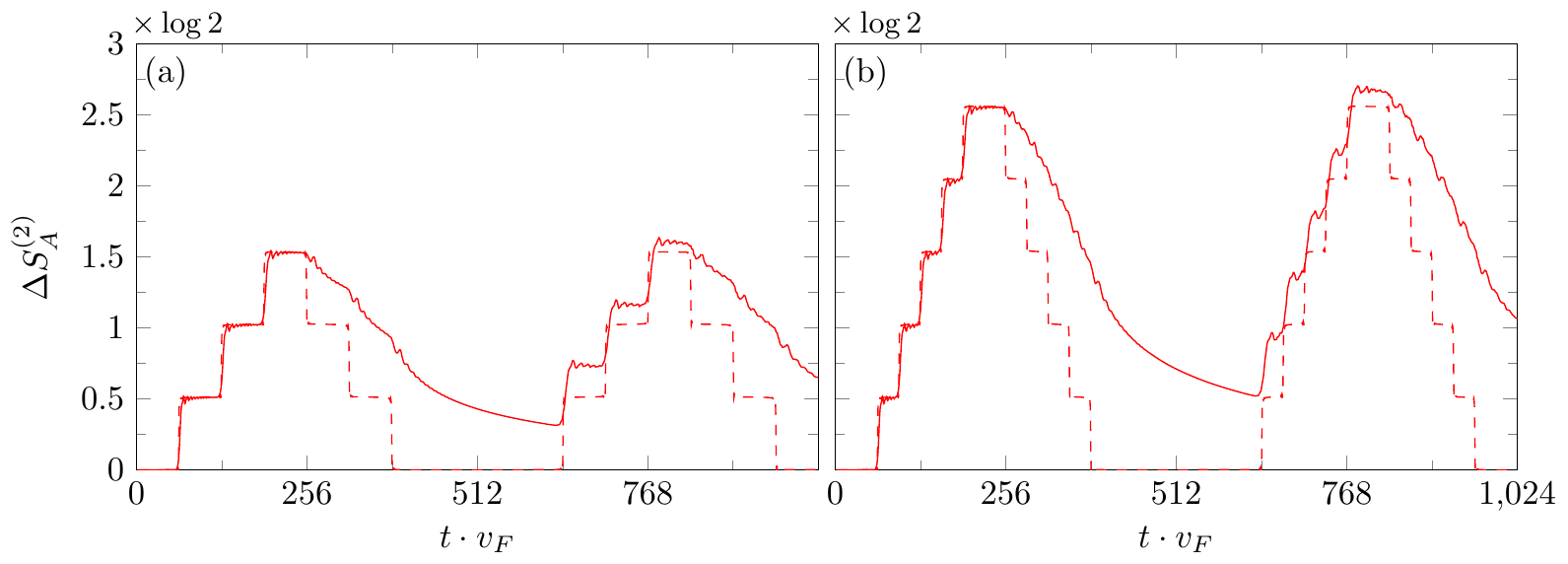}
\end{center}
\caption{Evolution after exciting by $\sigma(n)$ on (a) $3$ sites, $n \in \{1,65,129\}$ and (b) $5$ sites, $n \in \{1,33,65,97,129\}$. Results for Ising Hamiltonian (solid lines) and linearized Ising Hamiltonian (dashed lines).  Block $A$ supported on sites $193,\ldots,384$ ($L=192$).  $N=1024$. }
\label{steps}
\end{figure}

We begin with states where few operators were inserted on sufficiently separated sites and in some distance to the entangling block. We chose the block to be large enough so that there is a time where all the excited quasiparticles are inside the entangling region. From numerical results it is clear that in such excited states the entanglement entropy is a sum of the quantum dimensions of the operators, see Fig.~\ref{steps} for a case of 3 and 5 $\sigma$-excitations. The time evolution of entanglement in such states is very similar to periodic quench studied, e.g., in Ref.~\cite{PeschelEisler}.

Based on these observations, we conjecture that in rational CFTs, and more generally CFTs for which the quasi-particle picture remains a good effective description of the dynamics of entanglement, for $m$ such excitations the change in maximal contribution to the entanglement R\'enyi entropies is simply
\be
\Delta S^{(n)}_L=\sum^{m}_{i=1}\log d_i
\ee 
with quantum dimension of the $i$th operator $d_i$ (can be different or the same). We verify numerically, using $\hat H_{lin}$, that if we use some combination of operators $\sigma(n)$, $\epsilon(n)$ and $d\sigma(n)$ placed in the setup considered in Fig.~\ref{steps} where the insertion points are sufficiently separated, then the total increase of the entropy of the block $A$ indeed is a sum of contributions from single excitations discussed in section \ref{sec:num}. A similar result was reported in free quantum field theories in \cite{Nozaki:2014uaa} but it would be very interesting to formulate at least a necessary conditions for the validity of this formula in arbitrary interacting  2d CFT and we leave this as an open future problem.\\

Another class of excited states that we consider is defined by acting with local operators on all the sites of the chain
\be
\ket{\psi_G}=\prod^N_{i=1} \hat{ \mathcal{O}}_i\ket{0}.
\ee
They could be thought of as a version of a global quench \cite{Calabrese:2016xau} and have recently been employed in large $c$ holographic CFTs as states dual to the matter collapsing to a black hole \cite{Anous:2016kss}. 

\begin{figure}[t]
\begin{center}
\includegraphics[width=8cm]{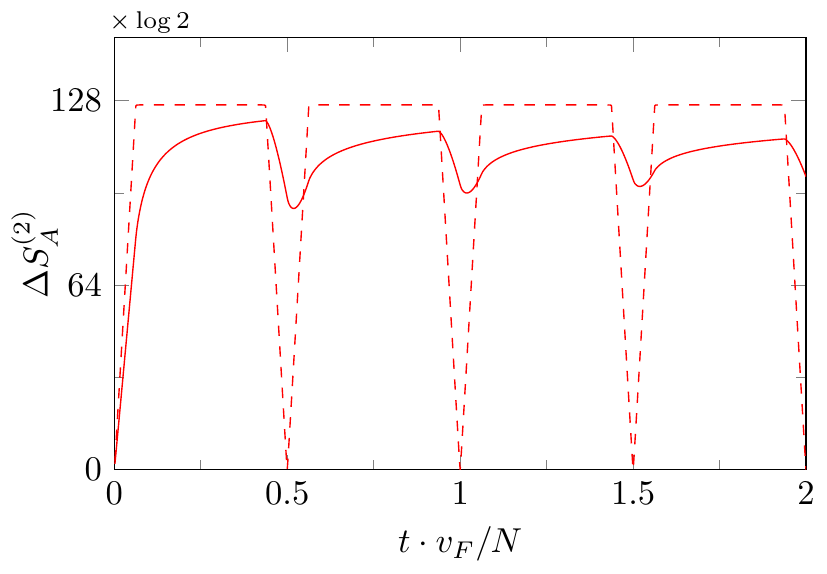}
\end{center}
\caption{Evolution after exciting all the sites by $\sigma^x$. Block size $L = 128$ and the system size $N=1024$. }
\label{Allsx}
\end{figure}

As we can see, the evolution looks qualitatively similar to the global quench in the finite size system \cite{Calabrese:2016xau}, but the final value in that case is given by the entropy density times the length of the interval. Here, the evolution of entanglement entropies can still be interpreted in terms of the quasiparticles propagating from the lattice sites to the left and right. This picture suggests that the maximal value of the entropies is equal to $\Delta S^{(n)}_{max}=2L\log d_O$, where the factor of two comes from the fact that at each site we can have two (left and right) quasiparticles. On the other hand, the maximal value of the entropy of the block of $L$ spins is attained by the density matrix with all $2^L$ eigenvalues equal and is also equal to $L \log 2 $. Apparently these two numbers coincide for $d_\sigma=\sqrt{2}$ and the value of the plateaux for $\Delta S^{(2)}$ is equal to this maximal value minus the ground state entropy, i.e. $S^{(2)}$ attains its maximal possible value, which can be seen in Fig.~\ref{Allsx} for $\hat H_{lin}$. If Ising Hamiltonian $\hat H$ is used instead, the maximal value of the plateaux is not reached and the clean periodic structure of revivals is obscured, showing the relevance of coherent dynamics of quasiparticles for obtaining those effects. In other words, the memory effects for the quasiparticles in the critical model seem to be suppressed by entanglement between quasiparticles with different momenta and for late times entanglement measures saturate. 

I would be interesting to compare how this behaviour changes for local operators with different quantum dimensions and for different densities of excitations leading possibly to a saturation, so we leave it as another open problem. Moreover, according to \cite{Anous:2016kss},  in the large $c$ CFTs,  entanglement entropy saturates at the thermal value with an effective temperature and the time for returning into the initial value (Poincare recurrence) is expected to be exponential in the central charge. It would be also interesting to further explore what happens in between these two regimes and how our memory effects are corrected once we consider states in chaotic or non-local toy models for black holes as for instance in Ref.~\cite{Mezei:2016wfz,Magan:2016ojb}.

\section{Relative entropy}  \label{sec:rel}
\label{sec:rel_ent}
We finish our numerical explorations with evolution of the relative entropy at criticality. As far as we are aware, numerical analysis of the relative entropy in critical systems has been much less explored than the R\'enyi entropies. Nevertheless, it is an important tool for understanding the notion of the distance between quantum states in field theories and plays an interesting role in uncovering the features of holographic CFTs \cite{Lashkari:2016idm}. The relative entropy is defined for two reduced density matrices $\hat \rho$ and $\hat \vartheta$ as
\be
S(\rho|\vartheta)=\Tr (\hat  \rho\log \hat  \rho )-\Tr (\hat  \rho\log \hat \vartheta ).
\label{eq:rel_ent}
\ee
In general excited state of a 2d CFT, the second term makes it very hard to compute analytically. Even for locally excited states the replica method requires the knowledge of the correlation function of $2n$ operators in order to continue to $n\to 1$, see, e.g., \cite{Lashkari:2015dia,Sarosi:2016oks}. On the other hand, if we compare excited states $\hat \rho$ with $\hat \vartheta$ obtained in the vacuum, given that the reduced density matrix $\hat \vartheta$ can be written as the exponent of a known modular Hamiltonian, $\hat \vartheta=e^{-\hat H_m}/Tr(e^{-\hat H_m})$, we can express the entropy as 
\be
S(\rho|\vartheta)=\Delta\langle \hat H_m\rangle-\Delta S^{(1)}.
\label{RRe}
\ee
The expectation value of the vacuum modular Hamiltonian $\Delta \langle \hat H_m\rangle=\Tr(\hat  \rho \hat H_m)-\Tr(\hat \vartheta \hat H_m)$ is computed in the excited state and $\Delta S^{(1)}$ denotes the difference of the von-Neumann entropies of the two density matrices, see \cite{Blanco:2013joa} for more details. 

Now, in 2d CFTs, the expectation value of the vacuum modular Hamiltonian for an interval of length $L$ in a state locally excited by a primary operator is universal. Namely, it follows from the OPE of the stress tensor with the primary operator, and can be computed from (see e.g. \cite{Cardy:2016fqc})
\be
\Delta\langle \hat H_m\rangle=\frac{\pi}{L}\int_Ldx (L-x)x\,\langle T_{00}(x)\rangle.
\ee
Moreover, as we argued in section \ref{sec:CFT}, the change of the entanglement entropy of the block is equal to the logarithm of the quantum dimension of the primary operator. From these two results we evaluate the relative entropy in our CFT setup and we can possibly compare it with numerics. 

\begin{figure}[t]
\begin{center}
\includegraphics[width=14cm]{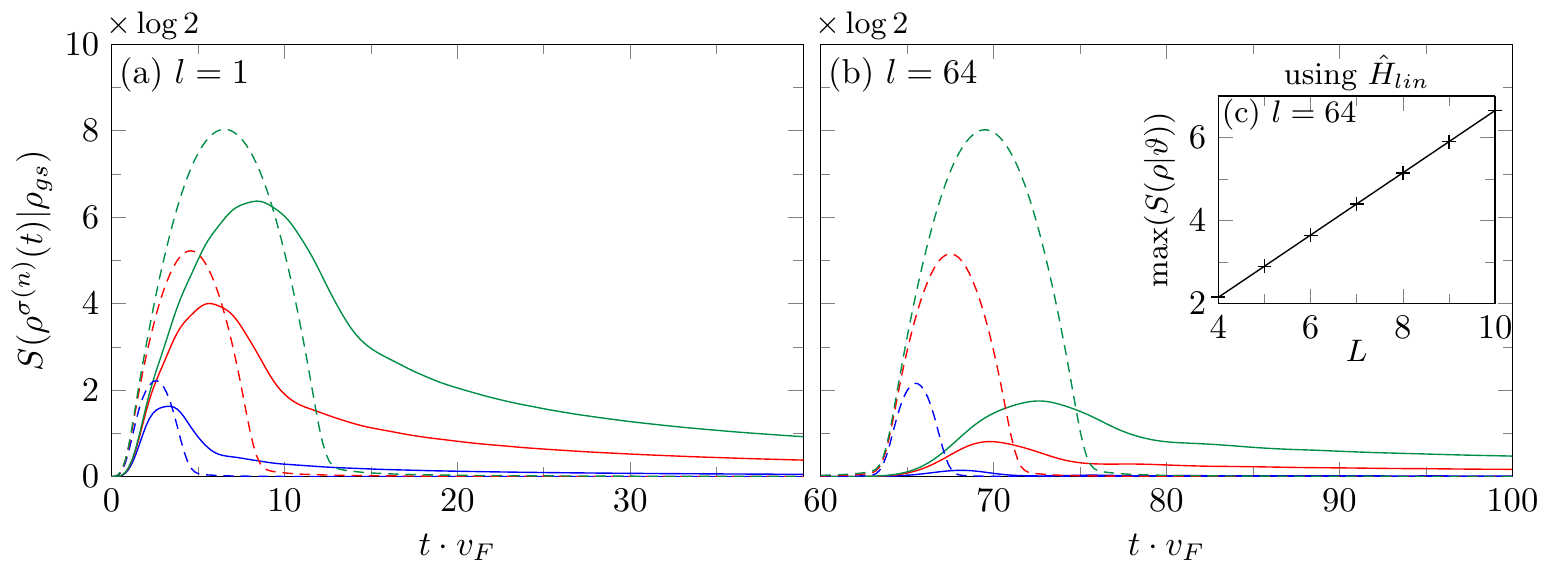}
\end{center}
\caption{Evolution of Relative Entropy after excitation by $\sigma(n)$. Colors indicate different block sizes $L=4$ (blue), $L=8$ (red), $L=12$ (green). Results for Ising Hamiltonian (solid lines) and linearized Ising Hamiltonian (dashed lines). Distance of the entangling block from the excitation (a) $l=1$, (b) $l=64$. $N=1024$. Inset (c) shows the maximum of relative entropy from panel (b) for linearized Ising Hamiltonian, where the linear fit $ \simeq 0.75 L - 0.85$. }
\label{fig:rel_entropy}
\end{figure}

In the following, we simply take $\hat \vartheta$ to be the reduced density matrix of a block of $L$ spins in the ground state and $\hat \rho$ as the density matrix \eqref{denmat} of the block for the state locally excited by operator $\sigma(n)$ in a distance $l$ from the block. We find an elegant expression for the relative entropy in terms of fermionic covariance matrices for Gaussian $\hat \rho$ and $\hat \vartheta$ 
\begin{equation}
S(\rho|\vartheta) = \Tr (C^\rho_L \log C^\rho_L) - \Tr (C^\rho_L \log C^\vartheta_L),
\end{equation}
and we refer to the Appendix \ref{Numer} for details. We only notice here that as $C^\rho_L$  and $C^\vartheta_L$, or equivalently $\hat \rho$ and $\hat \vartheta$, cannot be simultaneously diagonalized,  finite numerical precision limits the calculation of the second term in the above expression only to relatively small block sizes. We present the results in Fig. \ref{fig:rel_entropy}.

We observe that the signal strongly changes depending if $\hat H$ or $\hat H_{lin}$ is governing the time evolution. For the Ising Hamiltonian $\hat H$ the maximum of the relative entropy for given block size $L$  quickly disappears with the distance between the block and the excitation -- suggesting that modes with high momenta contribute significantly to the observed value. If linearized Ising Hamiltonian $\hat H_{lin}$ is used instead, the signal does not dissipate, with both the width and the value at the peak being proportional to $L$.

One can verify that, for non-zero $\epsilon$, the numerical evolution of the relative entropy is consistent with the CFT computation with the expectation value of the stress tensor in our locally excited state. Moreover, it is possible to compute in CFT the value of the peak of the relative entropy. The maximum of the relative entropy is universal in the small $\epsilon$ limit
\be
\max\left(S(\rho|\vartheta)\right) \simeq E_{\mathcal{O}}L+O(\epsilon),\label{FL}
\ee
where the energy due to the operator insertion is $E_\mathcal{O}\sim \Delta_{\mathcal{O}}/\epsilon$. Numerically, this is shown on the inset in Fig.~\ref{fig:rel_entropy} where we see the linear growth of the peak of the relative entropy with the length of the interval. 
Our numerical results in the Ising chain are consistent with the energy $E_{\sigma(n)} = \langle \Psi^{\sigma(n)} (t) | \hat H_{lin} / v_F | \Psi^{\sigma(n)} (t) \rangle - \langle 0| \hat H_{lin} / v_F | 0\rangle \simeq 0.742$ close to the slope $\simeq 0.75$ fitted in Fig.~\ref{fig:rel_entropy}(c).

Let us finally compare our formula \eqref{FL} with the first-law for entanglement entropy for a family of the nearby equilibrium states \cite{Blanco:2013joa,FLaw} that reads $\Delta E\sim T_e \Delta S$ with $T_e\sim L^{-1}$. This relation is a consequence of the vanishing relative entropy. Interestingly, we observe an analogous first-law like relation for the maximal value of the distance between our two quantum states as measured by the relative entropy. Moreover, the maximal value is universal and contained in $\Delta\langle \hat H_m\rangle$. The details of the CFT analysis for this class of locally excited states will appear elsewhere \cite{RelRen}.

\section{Discussion}
We have shown that, by using local operators on the critical lattice, one is able to extract further non-trivial CFT data -- like quantum dimensions -- numerically. We have successfully done so for the spin operator $\sigma$ and its derivative descendant. However, we also saw that, even after linearization of the dispersion relation, the lattice energy operator gave a non-zero contribution to the entropy of a block. This fact might be explained by the ambiguity in the identification between the lattice and the CFT operators or a non-trivial contribution from the ``tail" of operators in the continuum and deserves further investigation. 

Our analysis could be naturally extended to other CFTs with operators that enjoy known lattice counterparts. A natural setup might be the three-state Potts model in which \cite{Mong:2014ova} recently analyzed the lattice realization of the local operators in the parafermionic CFT. Clearly, it would be much more interesting to explore analytically what is the contribution from a general lattice operator and what kind of information can be extracted from the increase in R\'enyi entropies. In fact there has been a lot of work on extracting local CFT operators from the multiscale entanglement renormalization ansatz (MERA) \cite{EvVid}. It would be interesting to apply those developments in the studies of entanglement evolution after local operator excitations.

The evolution of entanglement measures after local excitations away from the critical point appears to be relatively unexplored. As we saw, the critical behavior appears to be very special -- characterized by formation of a clear plateaux -- and might be used as a smoking gun of a critical point. Moreover, the contribution to the R\'enyi entropies appears to be the smallest for the Ising Hamiltonian with critical parameters what might be a sign of a general bound. We hope that our analysis will serve as a starting point for more complicated systems and will help to uncover other unknown universal phenomena in the propagation of entanglement.

Last but not least, we performed the evolution of the relative entropy between the vacuum and a locally excited state. Interestingly, this distance measure shows universal features analogous to the first law of entanglement and its maximum -- maximal quantum distance -- is proportional to the change in the energy with an effective temperature. Exploring this relation in CFTs or free field theories where explicit computations are under control opens a new interesting path for investigation.
\\

{\bf Acknowledgements} 
We would like to thank John Cardy, Paul Fendley, Masahiro Nozaki, Tokiro Numasawa, Tadashi Takayanagi, Kento Watanabe, Luca Tagliacozzo and Alvaro Veliz-Osorio for correspondence and discussions. We would also like to thank Xueda Wen for comments sharing some of his unpublished results. We acknowledge support by the Swedish Research Council (VR) grant 2013-4329 (P.C.) and Narodowe Centrum Nauki (NCN, National Science Center) under Project No. 2013/09/B/ST3/01603 (M.M.R).  P.C. would like to thank Yukawa Institute for Theoretical Physics for hospitality and support during the "Quantum Information in String Theory and Many-body Systems" workshop where some of this work was performed. We would like to thank the organizers of the Tensor Network Summer School in Ghent where this work was initiated.

\appendix
\section{Details of simulations}\label{Numer}

{\it Hamiltonian.---} The Ising Hamiltonian \eqref{eq:Spin_Ising} is diagonalized in a standard way \cite{Lieb:1961} by mapping it onto a system of free fermions using the Jordan-Wigner transformation 
\begin{eqnarray}
\hat{\sigma}^z_n &=& 1-2 \hat{c}^\dagger_n \hat{c}_n,  \\
\hat{\sigma}^x_n + i \hat{\sigma}^y_n &=& 2 \hat{c}_n \prod_{m<n}(1-2 \hat{c}^\dagger _m \hat{c}_m), \nonumber
\end{eqnarray} 
where $\hat{c}_n$ are fermionic annihilation operators. For convenience, we introduce Majorana fermions 
$\hat{a}_{2n-1} = \hat{c}_n + \hat{c}^\dagger_n$, $\hat{a}_{2n} = i (\hat{c}_n-\hat{c}^\dagger_n)$ which are hermitian, unitary and satisfy canonical anticommutation relations $[\hat{a}_m , \hat{a}_n]_+ =  2 \delta_{m,n}$. The Ising Hamiltonian with periodic boundary conditions can be rewritten as
\begin{equation}
\hat{H} =   \hat{H}^{+}   \hat{P}^{+} +   \hat{H}^{-} \hat{P}^{-}.
\end{equation}
Above, $\hat{P}^\pm = \frac{1}{2} \left( 1 \pm \hat{P} \right)$ are projectors on the subspaces with respectively even and odd number of fermions, where the parity operator $\hat{P} = \prod_{n=1}^N \hat{\sigma}_n^z=e^{i \pi \sum_n \hat{c}_n^\dagger \hat{c}_n}$ commutes with $\hat{H}$. The Hamiltonians in both subspaces can be expressed in terms of fermionic operators as
\begin{equation}
\hat{H}^\pm = -\sum_{n=1}^{N} \left( \frac{i}{2} \hat{a}_{2n} \hat{a}_{2n+1} + \frac{i g}{2} \hat{a}_{2n-1} \hat{a}_{2n}+ h.c. \right), 
\end{equation}
differing only at the boundary term. This is accounted for by enforcing the boundary conditions: antiperiodic $\hat{a}_{2N+n} = -\hat{a}_{n}$ for $\hat{H}^{+}$, and periodic $\hat{a}_{2N+n} = \hat{a}_{n}$ for $\hat{H}^{-}$. For $g>0$, which we use in this work, for any value of $N$ the ground state of $\hat{H}$ belongs to the subspace with even parity, see e.g. Ref.~\cite{Damski:2014}.
 
In the following, we employ matrix notation to simplify description. It is convenient to rewrite $\hat{H}^{\pm} = {\vec{\hat{a}}}^\dagger {\rm H^\pm} \vec{\hat{a}}$, where $\vec{\hat{a}}$ is a column vector composed of operators $\{ \hat{a}_n \}$ and ${\rm H^\pm}$ are $2N \times 2N$ hermitian matrices. 
In each parity subspace the system is solved independently by a canonical transformation to new basis of Majorana fermions, respectively $\{ \hat d_n^+ \}$ and $\{ \hat d_n^- \}$ for $\hat H^+$ and $\hat H^-$, where
 $\vec{\hat{a}} = {\rm U}^{\pm} \vec{\hat{d}}^\pm$ and ${\rm U}^{\pm}$ are real and orthogonal matrices. The Hamiltonian then reads
\begin{equation}
\hat{H}^\pm=\sum_{k_n \in k^\pm} \epsilon^\pm_{k_n} \left({{\hat{\gamma }^{\pm \dagger }_{k_n}}}  \hat{\gamma}^\pm_{k_n} - \frac12\right),
\label{eq:Hpm}
\end{equation}
expressed above in terms of standard annihilation operators $\hat{\gamma}^\pm_{k_n} = \hat{\gamma}^\pm_{n} = \left(\hat{d}^\pm_{2n-1} -i \hat{d}^\pm_{2n} \right)/2$. This can be done analytically by a subsequent Fourier and Bogoliubov transformations \cite{Lieb:1961}, resulting in dispersion relation $\epsilon_k$ given by Eq.~\eqref{eq:epsilon_k}. The fermionic modes are indexed by quasi-momenta consistent with the respective boundary conditions, i.e.~ $k_n \in k^+ = \left\{ \pm \frac{\pi}N, \pm \frac{3 \pi}N, \pm \frac{5 \pi}N, \ldots \right\} \subset (-\pi,\pi]$ for $H^+$, and $k_n \in  k^- = \left\{ 0, \pm \frac{2 \pi}N, \pm \frac{4 \pi}N, \ldots\right\} \subset (-\pi,\pi] $ for $H^-$.  This procedure is equivalent to bringing ${\rm H}^\pm$ into the canonical form ${{\rm U}^{\pm}}^{ \dagger} {\rm H^\pm} {\rm U}^\pm = \bigoplus_{k_n \in k^{\pm}} 
\begin{pmatrix}
0 & - i \epsilon_{k_n}/2 \\
 i \epsilon_{k_n}/2  & 0 \\
\end{pmatrix}$. We can now formally introduce the linearized Ising Hamiltonian by matrices $ {\rm H}^\pm_{lin}$ defined as ${{\rm U}^{\pm}}^{ \dagger} {\rm H} ^\pm_{lin} {\rm U}^{\pm} =  \bigoplus_{k_n \in k^{\pm}} 
\begin{pmatrix}
0 & - i \epsilon^{lin}_{k_n}/2 \\
 i \epsilon^{lin}_{k_n}/2  & 0 \\
\end{pmatrix}$, where the linearized dispersion relation $\epsilon^{lin}_{k} = v_F |k| = 2 |k|$.

The ground state of the system,  $|0\rangle$, which is the starting point for our further considerations, is the even parity state annihilated by all annihilation operators diagonalizing $\hat{H}^+$, namely $\hat{\gamma}^+_{k_n} | 0 \rangle = 0$.
All the information about this state is encoded in matrix  ${\rm U}^+$.

{\it Local operators and time evolution.---}  
We consider local operators which in the fermionic language, up to irrelevant phase factors, read
\begin{eqnarray}
\sigma(n) = \hat{\sigma}_n^x &=& \prod_{m = 1}^{2n-1} \hat{a}_m,  \nonumber \\
\varepsilon(n) = \hat{\sigma}_n^x \hat{\sigma}_{n+1}^x - \sigma_n^z &=& \hat{a}_{2n} \left(\hat{a}_{2n-1} + \hat{a}_{2n+1} \right), \label{eq:app_operators} \\
d\sigma(n) = \hat{\sigma}_{n+1}^x - \hat{\sigma}_n^x  &=&  \left(\prod_{m = 1}^{2n} \hat{a}_m \right) \left( i \hat{a}_{2n+1} - \hat{a}_{2n} \right) \nonumber.
\end{eqnarray}
We simulate the action of those operators on the ground state by employing the Heisenberg picture. In order to explain our approach we consider operator $\hat Q = \hat q_l \ldots \hat q_2 \hat q_1$, where each $\hat q_j$ is a linear combination of Majorana fermions $\hat a_n$. All operators in Eq.~\eqref{eq:app_operators} are of such form, we note, however, that for general operator $\hat Q$  such simple decomposition usually does not exist.

Operators $\hat q_j$ above do not have to be unitary, so in order to move to the Heisenberg picture we find unitary operator
$\hat O = \hat o_l \ldots \hat o_2 \hat o_1$, being a product of unitary linear combination of  Majorana fermions $\hat a_n$,  such that
\begin{equation}
 \frac{\hat Q | 0\rangle} {||\hat Q | 0\rangle || } = \hat O | 0 \rangle. 
\end{equation}
We show how to construct such an operator below, but firstly, we present the discussion of calculations in the Heisenberg picture. We start with 
$\hat o_1 =  
\vec {\mathrm{v}}_1 \vec {\hat a}$, where $\vec {\mathrm{v}}_1$ is a row vector of coefficients. We remind that we employ notation where operators, e.g. Majorana fermions $\vec {\hat a}$,  form a column vector, coefficients would form a row vectors and multiplications should be understood as a standard matrix multiplication.
As $\hat o_1$ is unitary, $\vec {\mathrm{v}}_1$ is real (up to an irrelevant global phase factor which can be ignored) and normalized $\vec {\mathrm{v}}_1 \vec {\mathrm{v}}_1^\dagger = 1$. The operators in the Heisenberg picture read $\hat a_n^{o_1} =  \hat{o}_1^\dagger {\hat{a}_n} \hat{o}_1$, or in the matrix notation
\be
\vec{\hat{a}}^{o_1} = \hat{o}_1^\dagger \vec{\hat{a}} \hat{o}_1 =  \left(2 {\vec {\mathrm{v}}_1} ^\dagger  \vec{\mathrm{v}}_1  - \mathbb{1}\right)  \vec{\hat{a}} = \mathrm{P^1} \vec{\hat{a}}.
\ee
It is important to notice that $\vec{\hat{a}}^{o_1}$ is a linear combination of original Majorana fermions where $\mathrm{P^1}$ is the orthogonal transformation matrix.
In the following, in order to calculate entropies, we need the state $\hat o_1 |0\rangle$ -- and more generally $ e^{-i \hat H t} \hat O |0\rangle$ --  to be Gaussian. This can be seen by considering the expectation value $\langle 0 | \hat o_1^\dagger \hat a_{i_1} \hat a_{i_2} \hat a_{i_3} \ldots \hat o_1 | 0 \rangle = \langle 0 |  \hat a_{i_1}^{o_1} \hat a_{i_2}^{o_1} \hat a_{i_3}^{o_1} \ldots | 0 \rangle = \sum_{j_1 j_2 \ldots} P^1_{i_1 j_1} P^1_{i_2 j_2} P^1_{i_3 j_3} \ldots \langle 0 | \hat a_{j_1} \hat a_{j_2} \hat a_{j_3} \ldots   | 0 \rangle $. Now, as the state $ | 0 \rangle$  supports the Wick's theorem, so is $o_1 | 0 \rangle$ from its multi-linear character, and as such state $o_1 | 0 \rangle$ is Gaussian. This is also true for $\hat O | 0 \rangle$  by iterating this procedure. The reasoning extends, of course, to general unitary operator $\hat U$ such that $\hat U^\dagger  \vec{\hat{a}} \hat U = \mathrm{P} \vec{\hat{a}}$, for example $\hat U = e^{-i \hat H t}$ where $\hat H$ is free-fermionic Hamiltonian, as well as to the mixed Gaussian states.

Summarizing the above, the orthogonal matrix ${\rm U}^{(1)} =  \left(2 \vec {\mathrm{v}}_1^\dagger \vec {\mathrm{v} }_1 - \mathbb{1}\right)  {\rm U}^+ =  \mathrm{P^1} {\rm U}^+$ describes rotation  $\vec{\hat{a}}^{o_1} = {\rm U}^{(1)} \vec{\hat{d}}^+$ to the base in which $|0\rangle$ is the vacuum state.
Subsequent application of unitary operators $\hat{o}_j$ for $j=2,3,\ldots l$ is described recursively as $\hat{a}^{o_j}_n = (\hat{o}_j^{o_{j-1}})^\dagger a^{o_{j-1}}_n \hat{o}_j^{o_{j-1}} =  \left(2 \vec {\mathrm{v}}_j^\dagger \vec {\mathrm{v} }_j - \mathbb{1}\right) {\rm U}^{(j-1)} \vec{\hat{d}}^+ = {\rm U}^{(j)} \vec{\hat{d}}^+$, leading to  ${\rm U}^{(j)} = \left(2 \vec {\mathrm{v}}_j^\dagger \vec {\mathrm{v} }_j - \mathbb{1}\right)  {\rm U}^{(j-1)}$, where $\hat o_j =  
\vec {\mathrm{v}}_j \vec {\hat a}$. This finally results in $\vec{\hat{a}}^{o_l} =\vec{\hat{a}}^{O} =  \hat O^\dagger \vec{\hat{a}} \hat O = {\rm U}^{O}  \vec{\hat{d}}^+$ where for clarity we use ${\rm U}^{O}  = {\rm U}^{(l)}$.

The time evolution is simulated  in the Heisenberg picture as $\frac{\partial}{\partial t} \hat{a}^O_n(t) = i  [\hat{H}^O(t),\hat{a}^O_n(t)] $, which leads to
$ \vec{\hat{a}}^O(t) = {\rm U}^O(t) \vec{\hat{d}}^+$, where ${\rm U}^O(t) = e^{-4 i {\rm H} t} {\rm U}^{O}$ and the matrix ${\rm H} = {\rm H}^{+} ({\rm H}^{-}) $ for the even (odd)  parity of the excited state $\hat O |0\rangle$. Notice that operator $\epsilon(n)$ is preserving the parity, and $\sigma(n)$, $d\sigma(n)$ are changing it. Finally, the time evolution with the linearized Hamiltonian is simulated by using matrices ${\rm H}^{\pm}_{lin}$ instead of ${\rm H}^{\pm}$ in the expression above.

Finally, we address the case of general operator $\hat Q=\hat q_l \ldots \hat q_2 \hat q_1$.  We start with $\hat q_1 = \vec {\mathrm{w}}_1 \vec{\hat  a}$. If up to a normalization it is unitary, then $\hat o_1 = \hat q_1 / ||  \vec {\mathrm{w}}_1 ||$ and $\vec {\mathrm{v}}_1 =   \vec {\mathrm{w}}_1 /  ||  \vec {\mathrm{w}}_1 ||$. Let's consider the case when it is not unitary, i.e. the coefficients $\vec {\mathrm{w}}_1$ are not real. Since we are only interested in the action of $\hat q_1$ on the state $|0\rangle$, we express $\hat q_1 |0\rangle = \vec {\mathrm{w}}_1 \vec{ \hat  a} |0\rangle = \vec {\mathrm{w}}_1 {\rm U}^+ \vec d^+ |0\rangle = \sum_{n=1}^N \left( [ \vec {\mathrm{w}}_1 {\rm U}^+]_{2n-1} - i [ \vec {\mathrm{w}}_1 {\rm U}^+]_{2n} \right) \gamma^{+\dagger}_n |0\rangle$, where we used the fact that $|0\rangle$ is the vacuum state annihilated by annihilation operators $\gamma^{+}_n$. We can now define unitary operator $\hat o_1 = \vec {\mathrm{v}}_1 \vec {\hat a} = \frac{\vec {\mathrm{s}}_1 }{||\vec {\mathrm{s}}_1 ||}  \vec d^+ $, where $[\vec{ \mathrm{s}}_1]_{2n-1} = \Re\left([\vec {\mathrm{w}}_1 {\rm U}^+]_{2n-1} - i [ \vec {\mathrm{w}}_1 {\rm U}^+]_{2n}\right)$,
 $[\vec{ \mathrm{s}}_1]_{2n} = -\Im\left([\vec {\mathrm{w}}_1 {\rm U}^+]_{2n-1} - i [ \vec {\mathrm{w}}_1 {\rm U}^+]_{2n}\right)$ and $\vec {\mathrm{v}}_1  = \frac{\vec {\mathrm{s}}_1 }{||\vec {\mathrm{s}}_1 ||} { \mathrm{U^+}}^{\dagger}$, for which $\hat o _1 | 0 \rangle = \frac{\hat q _1 | 0 \rangle}{||\hat q _1 | 0 \rangle||}$. Clearly, such an operator always exists as long as $\hat q _1 | 0 \rangle \neq 0$.

The procedure can now be iterated to other $\hat q_j$. Notice that now we are interested in the action of $\hat q_j = \vec {\mathrm{w}}_j \vec {\hat a}$ on the state $\hat o_{j-1} \ldots \hat o_{2} \hat o_{1} |0\rangle$.
It is convenient to employ Heisenberg picture, where we require that $\hat o _j^{o_{j-1}} | 0 \rangle = \frac{\hat q _j^{o_{j-1}} | 0 \rangle}{||\hat q _j^{o_{j-1}} | 0 \rangle||}$. As above, this leads to $\hat o_j^{o_{j-1}} = \vec {\mathrm{v}}_j  \vec {\hat a}^{o_{j-1}} = \vec {\mathrm{v}}_j  \mathrm{U}^{(j-1)} \vec d^+  = \frac{\vec {\mathrm{s}}_j }{||\vec {\mathrm{s}}_j ||}  \vec d^+$, where $[\vec{ \mathrm{s}}_j]_{2n-1} = \Re\left([\vec {\mathrm{w}}_j {\rm U}^{(j-1)}]_{2n-1} - i [ \vec {\mathrm{w}}_j {\rm U}^{(j-1)}]_{2n}\right)$, $[\vec{ \mathrm{s}}_j]_{2n} = -\Im\left([\vec {\mathrm{w}}_j {\rm U}^{(j-1)}]_{2n-1} - i [ \vec {\mathrm{w}}_j {\rm U}^{(j-1)}]_{2n}\right)$ and  $\vec {\mathrm{v}}_j  = \frac{\vec {\mathrm{s}}_j }{||\vec {\mathrm{s}}_j ||} ({\rm U}^{(j-1)})^\dagger$.

{\it Calculating entropies.--- } The entropy of a block of consecutive spins is calculated in a standard way \cite{Peschel}. Since the state $ |\Psi^{O}(t) \rangle = e^{-i \hat H t} \hat O |0\rangle$ is Gaussian,  as argued above, all the information about the reduced density matrix of a block is encoded in the 
 $2L\times 2L$ covariance matrix 
\begin{equation}
C_L = \left[ \frac12{\langle \Psi^{O}(t) | \hat{a}_m \hat{a}_n  | \Psi^{O}(t) \rangle}\right]_{m,n = 2L_0+1,2L_0+2, \dotsc,2L_0+2L}
\end{equation}
supported on $L$ consecutive lattice sites $L_0+1, \dots, L_0+L$. It is found as a submatrix of $C = \frac12{\langle \Psi^O(t) | \vec{\hat{a}} \vec{\hat{a}}^\dagger  | \Psi^O(t) \rangle}= \frac12{\langle 0 | \vec{\hat{a}}^O(t) \vec{\hat{a}}^O(t)^\dagger  | 0 \rangle} =  \mathrm{U}^{O}(t)  C_v  \mathrm{U}^{O}(t)^{\dagger} $, with $C_v = \frac12 \langle0|\vec {\hat{d}}  \vec {\hat{d}}^\dagger |0\rangle= \bigoplus_{n =1}^N 
\begin{pmatrix}
1/2 & - i/2 \\
 i/2   & 1/2 \\
\end{pmatrix}$ being the correlation matrix in the canonical (vacuum) base.

R\'enyi  entropy of the block is then found as, 
\be
S^{(\alpha)}_L = \frac{1}{1-\alpha} \sum_{j=1}^L \log \left[p_j^\alpha + (1-p_j)^\alpha \right],
\ee
where $\{p_1, 1-p_1, p_2, 1-p_2, \ldots, p_L, 1-p_L \}$ are the eigenvalues of $C_L$. For  $\alpha = 1$, the von-Neumann entropy reads
\be
S^{(1)}_L = -  \sum_{j=1}^L \left[ p_j \log p_j + (1-p_j) \log (1-p_j) \right] = - \Tr (C_L \log C_L),
\ee
where the second equation provides the direct expression in term of covariance matrix $C_L$. Relative entropy can be similarly found  as
\begin{equation}
S(\rho|\vartheta)=\Tr (\hat  \rho\log \hat  \rho )-\Tr (\hat  \rho\log \hat \vartheta ) =\Tr (C^\rho_L \log C^\rho_L) - \Tr (C^\rho_L \log C^\vartheta_L),
\label{eq:app_Srel}
\end{equation}
where $\hat \rho$ and $\hat \theta$ are a reduced density matrices supported on the same consecutive $L$ lattice sites, respectively, in the states $|\Psi^{O}(t)\rangle$ and $|0\rangle$.

As the matrices $C^\rho_L = C_L$ and $C^\vartheta_L$ (calculated as the correlation matrix in the ground state) cannot be diagonalized simultaneously, it becomes difficult to numerically calculate the second term in the above equation for large blocks.
In that case $C^\vartheta_L$ has many eigenvalues approaching $0$, falling below numerical precision, which prevents the calculation of the logarithm. For this reason, in Sec.~\ref{sec:rel_ent}, we calculate residual entropy only for small enough block where this problem does not yet occur.

Finally, in order to derive the non-standard second term on the r.h.s. of Eq.~\eqref{eq:app_Srel} we employ the fact that $\hat \vartheta$ is Gaussian and can be expressed  
as $\hat \vartheta = \prod_{j=1}^L \left( q_j \hat f_j \hat f_j^\dagger +(1-q_j)  \hat f_j^\dagger \hat f_j \right)$, where the diagonal form in terms of fermionic annihilation (creation) operators $\hat f_j$ ($\hat f_j^\dagger$) can be directly obtained from the correlation matrix 
$C^\vartheta_L = \frac12 \langle 0|\vec{\hat  a}^L \vec{\hat  a}^{L \dagger}   |0\rangle$ \cite{Peschel}.  We introduce here $\vec{\hat a}^L$ as a column vector consisting of Majorana fermions $\hat a_{2L_0+1},\hat a_{2L_0+2}, \dotsc,\hat a_{2L_0+2L}$ in the entangling block. Then $\frac{1}{\sqrt{2}}\vec{\hat  a}^L =  \mathrm{U}^{\vartheta} \vec{\hat  f}$, where $\vec{\hat  f}$ is a column vector consisting of $\hat f_1, \hat f_1^\dagger, \hat f_2, \dotsc, f_L^\dagger$, and $\mathrm{U}^\vartheta$ is a unitary matrix diagonalising $C_L^\vartheta = \mathrm{U}^\vartheta \left[ \bigoplus_{j =1}^L 
\begin{pmatrix}
q_j & 0 \\
 0  & 1-q_j \\
\end{pmatrix} \right] \mathrm{U}^{\vartheta \dagger}$.
This allows to compute 
$\Tr \left( \hat \rho \log \hat \vartheta  \right) = 
\sum_{j=1}^L \langle f_j f_j^\dagger \rangle_{\hat \rho}  \log q_j  +  \langle f_j^\dagger f_j\rangle_{\hat \rho} \log (1-q_j) $. Using matrix notation this equals
$\Tr \left(  \left[ \bigoplus_{j =1}^L  \begin{pmatrix} \langle f_j f_j^\dagger \rangle_{\hat \rho}  & 0 \\  0  & \langle  f_j^\dagger f_j \rangle_{\hat \rho}\\\end{pmatrix}\right] \left[ \bigoplus_{j =1}^L \begin{pmatrix} \log q_j & 0 \\  0  & \log (1-q_j) \\ \end{pmatrix} \right]  \right)  = 
\Tr \left( \mathrm{U}^{\vartheta \dagger} C_L^\rho \mathrm{U}^{\vartheta}  \left[ \bigoplus_{j =1}^L  \begin{pmatrix} \log q_j & 0 \\ 0  & \log (1-q_j) \\ \end{pmatrix} \right] \right)  =  
\Tr \left( C^\rho_L \log C^\vartheta_L \right)$, which gives Eq.~\eqref{eq:app_Srel}.



\end{document}